\documentclass[final]{article}

\usepackage{corl_2020} 
\usepackage[cmex10]{amsmath}
\usepackage{amsfonts, amssymb, amsthm}
\usepackage{mathrsfs}
\usepackage[mathscr]{euscript}
\usepackage{graphicx}
\usepackage[numbers]{natbib}
\usepackage{multicol}
\usepackage{color}
\usepackage{floatrow}
\usepackage{booktabs}
\usepackage{wrapfig}
\usepackage{subcaption}
\usepackage{needspace}

\newcommand{\myparagraph}[1]{\needspace{1\baselineskip}\medskip\noindent {\bf #1}}

\input{mySymbol.sty}
\renewcommand{\bbR}{{\mathbb{R}}}

\newcommand{\eg}{e.g.,\ }
\newcommand{\ie}{i.e.,\ }

\newcommand{\comment}[1]{}

\def\Tr{\mathsf{T}}

\title{VGAI: End-to-End Learning of Vision-Based Decentralized Controllers for Robot Swarms}

\author{
  Ting-Kuei Hu\\
  Texas A\&M University\\
  \texttt{tkhu@tamu.edu} \\
  \And
  Fernando Gama \\
  University of Pennsylvania \\
  \texttt{fgama@seas.upenn.edu} \\
  \AND
  Tianlong Chen \\
  University of Texas at Austin\\
  \texttt{tianlong.chen@utexas.edu} \\
  
  \AND
  Zhangyang Wang \\
  University of Texas at Austin\\
  \texttt{atlaswang@utexas.edu} \\
  \And
  Alejandro Ribeiro \\
  University of Pennsylvania \\
  \texttt{aribeiro@seas.upenn.edu} \\
  \And
  Brian M. Sadler \\
  Army Resarch Laboratory \\
  \texttt{brian.m.sadler6.civ@mail.mil} \\
}

\begin{document}
\maketitle

\begin{abstract}
Decentralized coordination of a robot swarm requires addressing the tension between local perceptions and actions, and the accomplishment of a global objective. In this work, we propose to learn decentralized controllers based on solely raw visual inputs. For the first time, that integrates the learning of two key components:  \emph{communication} and \emph{visual perception}, in one \emph{end-to-end} framework. More specifically, we consider that each robot has access to a visual perception of the immediate surroundings, and communication capabilities to transmit and receive messages from other neighboring robots. Our proposed learning framework combines a convolutional neural network (CNN) for each robot to extract messages from the visual inputs, and a graph neural network (GNN) over the entire swarm to transmit, receive and process these messages in order to decide on actions. The use of a GNN and locally-run CNNs results naturally in a decentralized controller. We jointly train the CNNs and the GNN so that each robot learns to extract messages from the images that are adequate for the team as a whole. 
Our experiments demonstrate the proposed architecture in the problem of drone flocking and show its promising performance and scalability, \eg achieving successful decentralized flocking for large-sized swarms consisting of up to 75 drones.

\end{abstract}

\keywords{Vision-Based Control, Decentralized Control, CNNs, GNNs}

\section{Introduction}


Large-scale aerial swarms are being increasingly deployed for wireless networking, disaster response, among many other applications. 
Nowadays, most aerial robot swarms rely on the centralized control as a whole, either from a motion capture system (such as IMU sensory measurements) or global navigation satellite system (GNSS) \citep{position2011,Kushleyev2013,preiss2017,weinstein2018,vasarhelyiVSTSNV2014,varhelyiea2018}. These systems assume a central manager being able to access to global information at each time step \citep{global1987, global2003}, and that the decision of the whole group is made based on the optimal global policy. However, for scaling up to a large number of agents, centralized control is often challenged by its fragility to a single point of failures, or unreliable data links (even communication outage). Inspired by the collective motion of animal groups \citep{swarm2015}, decentralized control has been increasingly explored as a robust alternative for controlling large robot swarms. To make the whole collective decisions, a decentralized controller only involves locally perceived observations and local data exchanges in-between agents. 

Meanwhile, traditional swarm studies often assume the local agent observation from GNSS positions \citep{weinstein2018,position2011,Kushleyev2013,preiss2017,Tolstaya19-Flocking}. However, there are many practical situations in which such information is imprecise, unreliable or unavailable, e.g., when a small inter-drone distance is required. Recently, 
the visual modality has emerged to provide an unparalleled information density to maximize the autonomy for robotic systems 
\cite{zhu2018visdrone,wu2019delving}, as visual cameras are getting cheaper, lighter-weight, and more energy-efficient. Instead of assuming the available information of precise location or motion, each agent can visually perceive its immediate surroundings, to capture any change of location or velocity of another drone in its field of view (potentially long-range), with no delay caused by network propagation. Visual observations can also enable agents to gain sophisticated situational awareness, interpreting the semantically rich environment, and potentially executing more advanced swarm tasks.

Nevertheless, developing a decentralized control system based on local visual observations raise unique challenges to overcome. Unlike explicit IMU measurements of location or velocity that are directly related to control actions, visual input is harder to interpret, and can only implicitly infer each agent's own motion. 
The semantic gap between raw visual perception and end decision-making remains under-explored in the research community. Despite that in \citep{schilling2018learning} the authors pioneered the learning of an end-to-end mapping from the raw visual inputs to end actions, their framework was only demonstrated on small-scale swarms (9 agents). Their lack of information exchanges between nearby agents make the learned policy challenging to scale up. Secondly, transmitting visual inputs among agents, in the form of either raw visual images or extracted intermediate features, can often cause prohibitive bandwidth load and latency for wireless channels, calling for compact designs.

\begin{figure*}[ht]
    \centering
    \vspace{-0.1em}
    \includegraphics[width=0.9\linewidth]{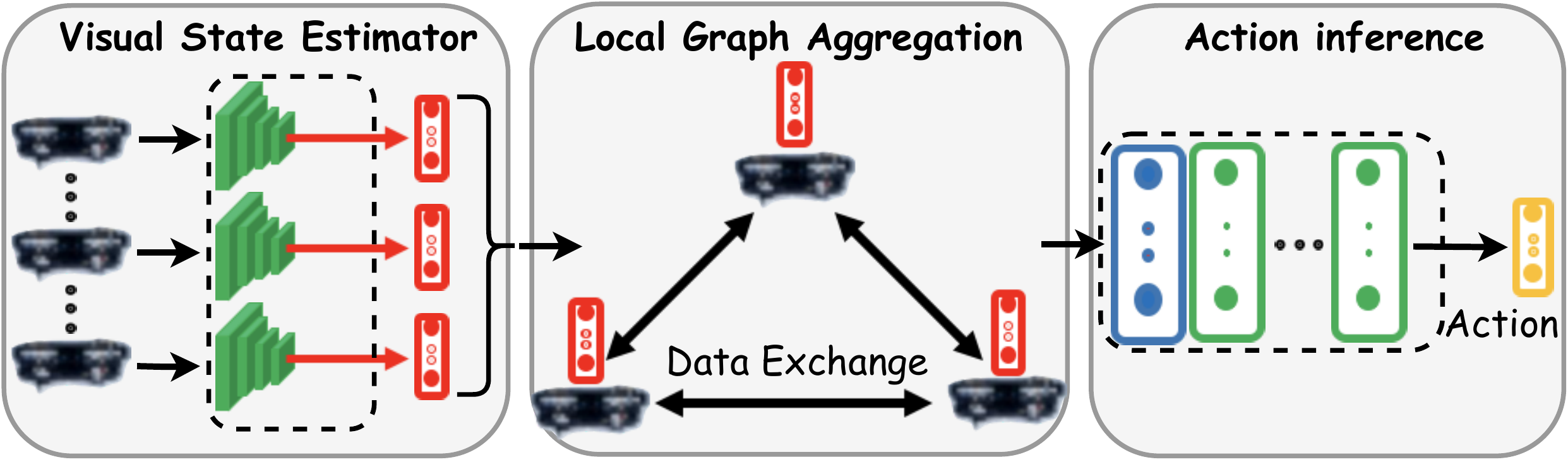}
   \vspace{-0.2em}
    \caption{The VGAI framework overview. Each agent's raw visual observation is first mapped into a compact visual state feature (Stage (i)). The neighbors then communicate to exchange only those compact visual state features and each agents aggregates the received features with its own (Stage (ii)). Eventually, the agent predicts its next action from the aggregated features (Stage (iii)).}
    \label{fig:VGAI}
   \vspace{-0.45em}
\end{figure*}


This work proposes \textit{Vision-based Graph Aggregation and Inference} (\textbf{VGAI}), the first decentralized learning-to-control framework that can (1) directly map raw visual observations to agent actions; and (2) scale up to large swarms owing to incorporating local communication. As illustrated in Fig. \ref{fig:VGAI}, VGAI consists of three stages: \textbf{(i)} visual state estimation; \textbf{(ii)} local graph aggregation; and \textbf{(iii)} action inference. Stages (i) and (iii) are by each agent individually, while Stage (ii) involves local sparse information aggregation. That is implemented by a cascade of agent-level convolutional neural networks (CNNs) 
for Stage (i), and a swarm-level graph neural network (GNN) addressing Stages (ii) (iii), respectively. First, each agent has a CNN to process its visual input and map that into a compact \textit{visual state} features, for efficient transmission. Next, we refer to a recently proposed decentralized learning framework, called \textit{Delayed Aggregation Graph Neural Network} (DAGNN) \citep{Gama_2019, Gama19-Aggregation, Tolstaya19-Flocking}, where each agent (as a node) will fuse the received visual state features with its own, based on which the next action of this agent is predicted. The communication here is completely local (it requires, at most, repeated exchanges with the one-hop neighbors only). Eventually, the entire framework is \textit{jointly trained} from end to end.

To our best knowledge, VGAI is the first to integrate \textit{communication} with \textit{visual observations} for end-to-end \textit{decentralized controller} learning. Prior works either learn only communication (no visual learning) \cite{Tolstaya19-Flocking}, or involve only vision-based action inference (no communication) \citep{schilling2018learning}. In contrast, VGAI seamlessly combines the learning of \textit{agent visual perception} and \textit{local sparse communication}, and trains them jointly. That contributes to scaling up VGAI to medium- and large-sized swarms, e.g., up to \textbf{75 agents}. We examine the proposed framework on the application of flocking \citep{Tolstaya19-Flocking}. Experiments demonstrate the promising performance and scalability of our learned controller, that can perform even comparably to the centralized controller learned from global information.


\ifx
A swarm operates in a completely decentralized and self-organized manner: each agent is responsible for its own decision making, based on its own observations as well as purely local interactions with nearby agent. Such a decentralization property has shown superior robustness to single-point failures, a high scalability since the number of agents can be flexibly changed, and saving of communication bandwidth. 
Despite the benefits from decentralized control, the challenges arise from the unavoidable restriction of local information network structures, compared to selecting optimal control actions with global information access.  While explicit multi-hop message passing across the team is feasible for offline \citep{swarm2013} or incremental training \citep{robots2012}, it will incur a superlinear growth in communications with team size.

A recent work \citep{AGNN2019} has proposed to learn local controllers which require only local information and local communications at testing time, by imitating the policy of centralized controllers using global information at training time (which can have different and time-varying network structures from testing). The authors take advantage of drone sensor measurements such as position and velocity to estimate the states. The key building block, the aggregation graph neural networks (GNNs) \citep{Gama_2019}, are designed to operate on network data in a completely local and decentralized way, by repeated exchanges with neighbors. The authors of \cite{AGNN2019} further extended aggregation GNNs to Delayed Aggregation GNNs (DAGNNs), that could handle time-varying graph processes. Based on those, they use imitation learning \cite{mobilerobots2015} to train distributed policies while attempting to mimic the global policy of a clairvoyant central agent.

While the usage of basic sensor measurements (IMUs, etc.) has been popular in control and swarm studies, in natural biological swarms such as flocks of birds, animals often rely most on their visual perception. As the visual cameras are getting cheaper, lighter-weight, and lower in energy consumption, the visual modality has shown the potential to provide an unparalleled information density to maximize the autonomy for robotic systems, which seems to be further enabled by the recent progress in computer vision and deep learning. DroneNet \citep{dronet2018} pioneers to predict a steering angle and a collision probability for single drone collision avoidance based on visual inputs, by using a convolutional neural network (CNN) trained on collected labeled images. Authors in \citep{schilling2018learning} then takes the first step towards decentralized vision-based flocking. It generates 3D velocity commands directly from raw camera images using a CNN. The vision-based controller is shown to learn not only robust collision avoidance but also the coherence of the flock in a sample-efficient manner. The visualizations also demonstrate that CNN learns to localize other agents in the visual input without explicit supervision. However, transmitting visual inputs among drones, in the form of either raw data or extracted features, can cause prohibitive congestion and latency for wireless channels. Therefore, each agent in \cite{schilling2018learning} takes only each drone’s visual observation into account for control, without considering any form of communication. 

In view of the above progress, we take a further step by integrating visual inputs into DAGNN-based decentralized controller learning with local sparse communications. Instead of transmitting raw visual inputs at the expense of expensive communication costs, we utilized the visual inputs to approximate the non-linear state estimation for sparse communications. To the best of our knowledge, this is the first successful attempt to learn vision-based swarm behaviors directly from raw images with spare local communications to achieve collision-free flocking behaviours.
\fi

\vspace{-0.5em}
\section{Related Work}

\vspace{-0.5em}

It has long been known that finding optimal controllers in the decentralized settings is challenging \cite{witsenhausen1968counterexample}, due to the restriction of purely local interactions with nearby agents. Recent efforts on decentralized drone flocking algorithms have been made on designing local controllers which incorporate local observation from spatial neighbours \cite{global2003,global1987,local2003}. Particularly, a recent work \citep{Tolstaya19-Flocking} presents important progress based on information exchanges between multi-hop neighbors.
Their commonality is the requirement to access global navigation satellite system (GNSS) positions through wireless communication among flock members. However, there are many situations in which GNSS positions are imprecise, particularly in scenarios that require a small inter-drone distance. For example, in urban environments, tall buildings may deflect the GNSS signal causing imprecise position information.

\vspace{-0.5em}
\paragraph{Vision-based Single Drone Control}\label{sec:2-2}
Imitation learning is commonly used in vision-based drone control to design meaningful actions. The authors in \citep{mav2012} trained a controller that can avoid trees in the forest by adapting the MAVs heading. Extracted visual features are mapped to the control input provided by the expert. In the problem of single drone collision avoidance, DroneNet \citep{dronet2018} pioneers to predict a steering angle and a collision probability based solely on visual inputs, by formulating angle prediction as a regression problem and using a CNN trained on collected labeled images. 
Another approach based on reinforcement learning \citep{rf2016} shows that a neural network trained entirely in a simulated environment can generalize to real-world navigation and leads to rare collisions. Other data-driven approaches \citep{vision2016, vision2017, vision2017_2} have also shown generality to fly a robot in real-world environments. However, the aforementioned approaches are for operating and navigating a single drone, and do not extend to coordinating a large multi-agent swarm.

\vspace{-0.5em}
\paragraph{Vision-based Decentralized Control}\label{sec:2-3}
Early works attempted for decentralized vision-based drone control, by mounting special visual markers on the drones \citep{marker2013, marker2014}. However, these markers are often unrealistically large or heavy and therefore impractical for real-world deployment.

The authors in \citep{schilling2018learning} took the first step towards general decentralized vision-based flocking. In that work, each agent independently generates 3D velocity commands directly from solely raw camera images, using a CNN. The vision-based controller is shown to learn not only robust collision avoidance but also the coherence of the flock in a sample-efficient manner. 
Their framework and our proposed VGAI share the same goal, while a key difference lies that the former did not exploit local communication between nearby drone agents. Due the overall task complexity, the learned policy in \citep{schilling2018learning} was demonstrated to operate a small swarm of \textbf{9 agents}.



\section{Approach}


\subsection{Problem Setting: Decentralized Flocking}

Consider a set of $N$ agents $\ccalV = \{1,\ldots,N\}$. At time $t \in \mbN_{0}$, each agent $i \in \ccalV$ is described\footnote{For the sake of simplicity and without loss of generality, we set all agents to fly on the same height plane in the simulation software, therefore reducing the position and velocity to 2D vectors.} by its position $\bbr_{i}(t) = [r_{i}^{x}(t), r_{i}^{y}(t)]^{\Tr} \in \reals^{2}$, velocity $\bbv_{i}(t) = [v_{i}^{x}(t),v_{i}^{y}(t)]^{\Tr} \in \reals^{2}$ and acceleration $\bbu_{i}(t) = [u_{i}^{x}(t),u_{i}^{y}(t)]^{\Tr} \in \reals^{2}$. We consider $t$ to be a discrete-time index representing consecutive time sampling instances with interval $T_{s}$. 
%
The objective of flocking is to coordinate the velocities $\bbv_{i}(t)$ of all agents to be the same
\begin{equation} \label{eqn:flockingObjective}
    \min_{\substack{\bbu_{i}(t)\\ i=1,\ldots,N\\t \geq 0}} \frac{1}{N} \sum_{t} \sum_{i=1}^{N} \Big\| \bbv_{i}(t) - \frac{1}{N} \sum_{j=1}^{N} \bbv_{j}(t) \Big\|^{2}.
\end{equation}
%
Each agent can control its acceleration $\bbu_{i}(t)$ so that the resulting velocities $\bbv_{i}(t)$ satisfy \eqref{eqn:flockingObjective}. A solution to \eqref{eqn:flockingObjective} that avoids collisions is given by accelerations $\bbu_{i}^{\ast}(t)$ computed as \cite[eq. (10)]{Tolstaya19-Flocking}
\begin{equation} \label{eqn:optimalSolution}
    \bbu_{i}^{\ast}(t) = -\sum_{j=1}^{N} \Big( \bbv_{i}(t)-\bbv_{j}(t) \Big) - \sum_{j=1}^{N} \nabla_{\bbr_{i}(t)} U \Big( \bbr_{i}(t),\bbr_{j}(t) \Big)
\end{equation}
where
\begin{equation} \label{eqn:collisionAvoidance}
 U(\bbr_{i}(t),\bbr_{j}(t)) 
 =
        \begin{cases} 
            1/\|\bbr_{ij}(t)\|^{2} - \log(\|\bbr_{ij}(t)\|^{2}) & \text{if } \|\bbr_{ij}(t)\| \leq \rho \\
            1/\rho^{2} - \log(\rho^{2}) & \text{otherwise}
        \end{cases}
\end{equation}
is a collision avoidance potential, with $\bbr_{ij}(t) = \bbr_{i}(t) - \bbr_{j}(t)$ and $\rho$ the value of the minimum distance allowed between agents. It is evident that, in computing the optimal solution \eqref{eqn:optimalSolution}, each agent $i$ requires knowledge of the velocities of all other agents in the network. Thus, the optimal solution $\bbu^{\ast}(t)$ is a \emph{centralized} controller.

Our objective, in contrast, is to obtain a \emph{decentralized} solution that can be computed only with information perceived by each agent, in combination with information relied by neighboring agents. We determine that agents $i$ and $j$ are able to communicate with each other at time $t$ if $\|\bbr_{i}(t)-\bbr_{j}(t)\| \leq R$ for some given communication radius $R$. We describe the communication network by means of a succession of graphs $\ccalG(t) = \{\ccalV, \ccalE(t)\}$ where $\ccalV$ is the set of agents, and $\ccalE(t) \subseteq \ccalV \times \ccalV$ is the set of edges, i.e. $(i,j) \in \ccalE(t)$ if and only if $\|\bbr_{i}(t) - \bbr_{j}(t)\| \leq R$. The communication link $(i,j) \in \ccalE(t)$ allows for exchange of information between nodes $i$ and $j$ at time $t$. Denote by $\ccalN_{i}(t) = \{j \in \ccalV : (j,i) \in \ccalE(t)\}$ the set of all agents that can communicate with node $i$ at time $t$.

A possible \emph{heuristic} to obtain a decentralized solution is to compute \eqref{eqn:optimalSolution} considering only neighboring information, namely
\begin{equation} \label{eqn:localHeuristic}
    \tbu_{i}(t) = -\sum_{j\in \ccalN_{i}(t)} \Big( \bbv_{i}(t)-\bbv_{j}(t) \Big) 
     - \sum_{j\in \ccalN_{i}(t)} \nabla_{\bbr_{i}(t)} U \Big( \bbr_{i}(t),\bbr_{j}(t) \Big). \nonumber
\end{equation}
We note that solution \eqref{eqn:localHeuristic} relies only on present, one-hop information. We can thus improve on this solution by incorporating information from neighbors that are farther away. However, in order to maintain the decentralized nature of the solution, the information has to be relied across robots, getting inevitably delayed. In what follows, we propose to \emph{learn} a decentralized solution that incorporates delayed information from farther away neighborhoods.

\subsection{Local Graph Aggregation and Action Inference}

Let $\bbx_{i}(t) \in \reals^{F}$ be the state of agent $i$ at time $t$, described by an $F$-dimensional vector of \emph{features}. Denote by $\bbX(t) \in \reals^{N \times F}$ the row-wise collection of the state of all agents
\begin{equation} \label{eqn:stateMatrix}
    \bbX(t) = \begin{bmatrix} \bbx_{1}^{\Tr}(t) \\ \vdots \\ \bbx_{N}^{\Tr}(t)
    \end{bmatrix}.
\end{equation}
To describe the communication between agents, we define the graph shift operator (GSO) matrix $\bbS(t) \in \reals^{N \times N}$ which respects the sparsity of the graph, i.e. $[\bbS(t)]_{ij} = s_{ij}(t)$ is nonzero if and only if $(j,i) \in \ccalE(t)$. Examples of GSO used in the literature are the adjacency matrix \citep{Sandryhaila13-DSPG}, the Laplacian matrix \citep{Shuman13-SPG}, or respective normalizations \citep{Ortega18-GSP}. Due to the sparsity of the GSO $\bbS(t)$, right-multiplication of $\bbS(t)$ with $\bbX(t)$ can be computed only by means of local exchanges with neighboring nodes only, yielding
\begin{equation} \label{eqn:graphShift}
    [\bbS(t) \bbX(t)]_{if} = \sum_{j \in \ccalN_{i}(t)} s_{ij}(t) [\bbx_{j}(t)]_{f}
\end{equation}
for each feature $f=1,\ldots,F$. In essence, multiplication \eqref{eqn:graphShift} updates the state at each agent by means of a linear combination of the states of neighboring agents.

We build the \emph{aggregation sequence} \citep{Gama19-Aggregation}, gathering information from further away neighborhoods by means of $(K-1)$ repeated exchanges with our one-hop neighbors
\begin{equation} \label{eqn:aggregationSequence}
    \bbZ(t) = \big[ 
            \bbX(t), \
             \bbS(t) \bbX(t-1),  \
            \ldots, \
         \bbS(t) \cdots \bbS(t-(K-2)) \bbX(t-(K-1))
    \big].
\end{equation}
The aggregation sequence $\bbZ(t)$ is a $N \times KF$ matrix, where each $N \times F$ block $\bbZ_{k}(t)$ represents the \emph{delayed aggregation} of the state information at the neighbors located at $k$-hops. Denote by $\bbz_{i}(t) \in \reals^{FK}$ the row $i$ of matrix $\bbZ(t)$, which represents the information gathered at node $i$. We note that this information has been obtained by executing $(K-1)$ communication exchanges with one-hop neighbors, in an entirely local manner.

Once we have the collected neighboring information at each node, we can proceed to apply a neural network \citep{Goodfellow16-DeepLearning} on vector $\bbz_{i}(t)$ to map the local graph information into an action
\begin{equation} \label{eqn:NNlayer}
    \bbz_{\ell} = \sigma_{\ell} \big(\bbtheta_{\ell} \bbz_{\ell-1} \big) \ , \ \bbz_{0} = \bbz_{i}(t) \ , \ \bbu_{i}(t) = \bbz_{L}
\end{equation}
where $\bbz_{\ell} \in \reals^{F_{\ell}}$ represents the output of layer $\ell$, $\sigma_{\ell}$ is a pointwise nonlinearity (known as \emph{activation function}) and $\bbtheta_{\ell} \in \reals^{F_{\ell} \times F_{\ell-1}}$ are the learnable parameters. The input to the neural network is the aggregated sequence, $\bbz_{0} = \bbz_{i}(t)$, with $F_{0} = KF$, and we collect the resulting action as the output of the last layer $\bbu_{i}(t) = \bbz_{L}$, so that $F_{L} = 2$. We compactly describe the neural network as
\begin{equation} \label{eqn:NN}
    \hbu_{i}(t) = \text{NN}_{\bbTheta}\big(\bbz_{i}(t) \big)
\end{equation}
where $\bbTheta = \{\bbtheta_{\ell}, \ell=1,\ldots,L\}$ are the \emph{learnable} parameters of each layer.

Several important observations are in order. First, the neural network parameters $\bbTheta$ do not depend on the specific node $i$, nor on the specific time-index $t$. This is a weight sharing scheme that allows for scalability (i.e., once trained, it can be deployed on any number of agents), and prevents overfitting (i.e., it avoids a number of parameters that grows with the data dimension). Second, since the aggregation sequence has already incorporated the graph information [cf. \eqref{eqn:aggregationSequence}], applying a regular neural network to $\bbz_{i}(t)$ is already taking into account the underlying graph support, leading to an \emph{aggregation neural network} architecture \citep{Gama_2019, Gama19-Aggregation}. Third, the resulting architecture is entirely \emph{local} in the sense that, at test time, it can be implemented entirely by means of repeated communication exchanges with one-hop neighboring nodes only. This is seen in \eqref{eqn:aggregationSequence}, which states that each node receives messages from their neighbors, adds them up, and stores them as the corresponding row of the first block $\bbZ_{0}(t)$; then, receives new messages from their neighbors, adds them up, and stores them as the corresponding row of the second block $\bbZ_{1}(t)$, and so on. For each of the $K-1$ communication exchanges, each robot adds up the messages obtained from immediate neighbors and stores it in the aggregation sequence \eqref{eqn:aggregationSequence}.

To train the neural network \eqref{eqn:NN} we use \emph{imitation learning} \cite{Ross10-ImitationLearning}. That is, we assume availability of a training set comprised of trajectories $\ccalT = \{(\bbX(t), \bbU^{\ast}(t))_{t}\}$ where $\bbX(t)$ is the collection of states \eqref{eqn:stateMatrix} and $\bbU^{\ast}(t) \in \reals^{N \times 2}$ is the collection of optimal actions for each agent
\begin{equation} \label{eqn:actionMatrix}
    \bbU^{\ast}(t) = \begin{bmatrix} \bbu_{1}^{\ast}(t)^{\Tr} \\ \vdots \\ \bbu_{N}^{\ast}(t)^{\Tr}
    \end{bmatrix}.
\end{equation}
where $\bbu_{i}^{\ast}(t) \in \reals^{2}$ is the optimal action of agent $i$ at time $t$ given by the optimal controller \eqref{eqn:optimalSolution}. Then, the optimal parameters can be found as
\begin{equation} \label{eqn:imitationLearning}
    \bbTheta^{\ast} = \argmin_{\bbTheta} \sum_{\ccalT} \sum_{i=1}^{N} \| \hbu_{i}(t) - \bbu_{i}^{\ast}(t) \|
\end{equation}
with $\hbu_{i}(t) = \text{NN}_{\bbTheta}(\bbz_{i}(t))$ and $\bbz_{i}(t)$ row $i$ of the aggregation sequence built as in \eqref{eqn:aggregationSequence}.

In what follows, we consider the local state $\bbx_{i}(t)$ to be the output of the visual state estimator $\text{CNN}_{\bbPsi}(\cdot)$ described next. Note that $\bbz_{i}(t)$ depends on the input $\bbx_{i}(t)$ through the DAGNN \eqref{eqn:NN} which, in turn, depends on the input image through the visual state estimator $\text{CNN}_{\bbPsi}(\cdot)$. Thus, solving problem \eqref{eqn:imitationLearning} actually demands joint training of the GNN and the CNN. Finally, we observe that using imitation learning gives the framework the flexibility to consider solutions other than \eqref{eqn:optimalSolution}, for instance, incorporating a maximum distance to penalize agents wandering off.

\begin{figure}[ht]
    \centering
    \vspace{-0.5em}
    \includegraphics[width=1\linewidth]{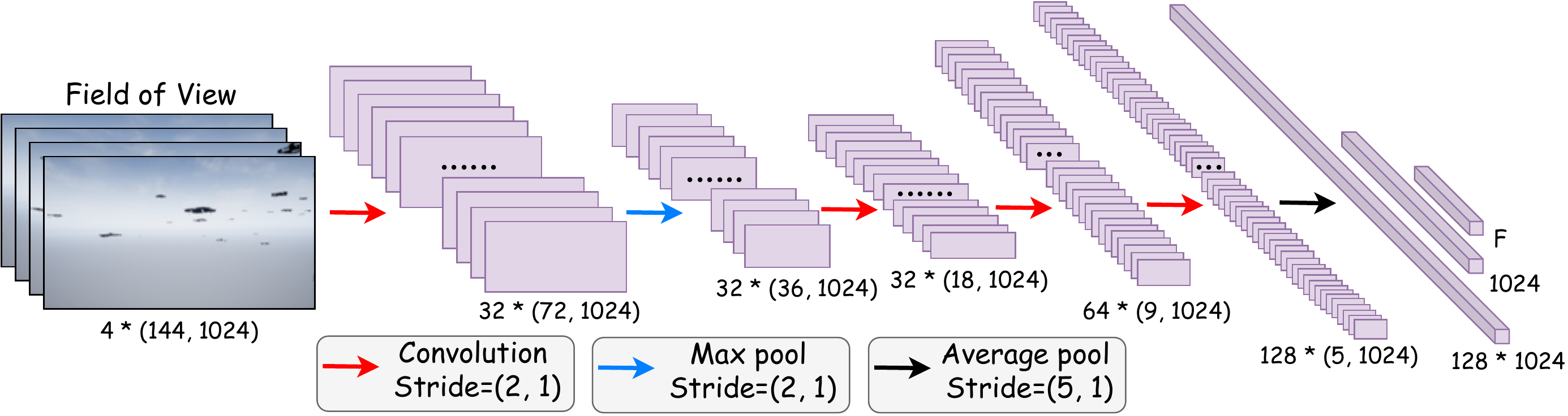}
    \caption{The architecture of our visual state estimator $\text{CNN}_{\bbPsi}(\cdot)$, consisting of four residual blocks and two fully-connected layers. The last layer's output, a $F$-dimensional vector, is the extracted visual state feature. One task-specific modification is that we only execute (average) pooling along the vertical axis while keeping the horizontal resolution intact, due to all our agents flying on the same height.}
    \label{fig:uni_droNet}
    \vspace{-0.5em}
\end{figure}

\subsection{Visual State Estimator}

As shown in Figure~\ref{fig:uni_droNet}, the objective of the visual state estimator is to extract \textit{compact} features  $\bbX(t)$ from the raw visual observation $\bbH(t)$ of a local agent, that can indicate its motion state and feed DAGNN \eqref{eqn:NN} for deciding the next action to take. The visual state feature $\bbX(t) \in \mathbb{R}^F$ is obtained from a CNN denoted as $\text{CNN}_{\bbPsi}(\cdot)$: $\bbX(t) = \text{CNN}_{\bbPsi}\big( \bbH(t) \big)$, where $\bbPsi$ is the set of learnable weights. The output feature dimension $F$ can be chosen as a small number for efficient transmission, and its ablation study can be found in Section 4.1



\begin{wrapfigure}{r}{80mm}
\vspace{-4em}
    \centering
    \includegraphics[width=1\linewidth]{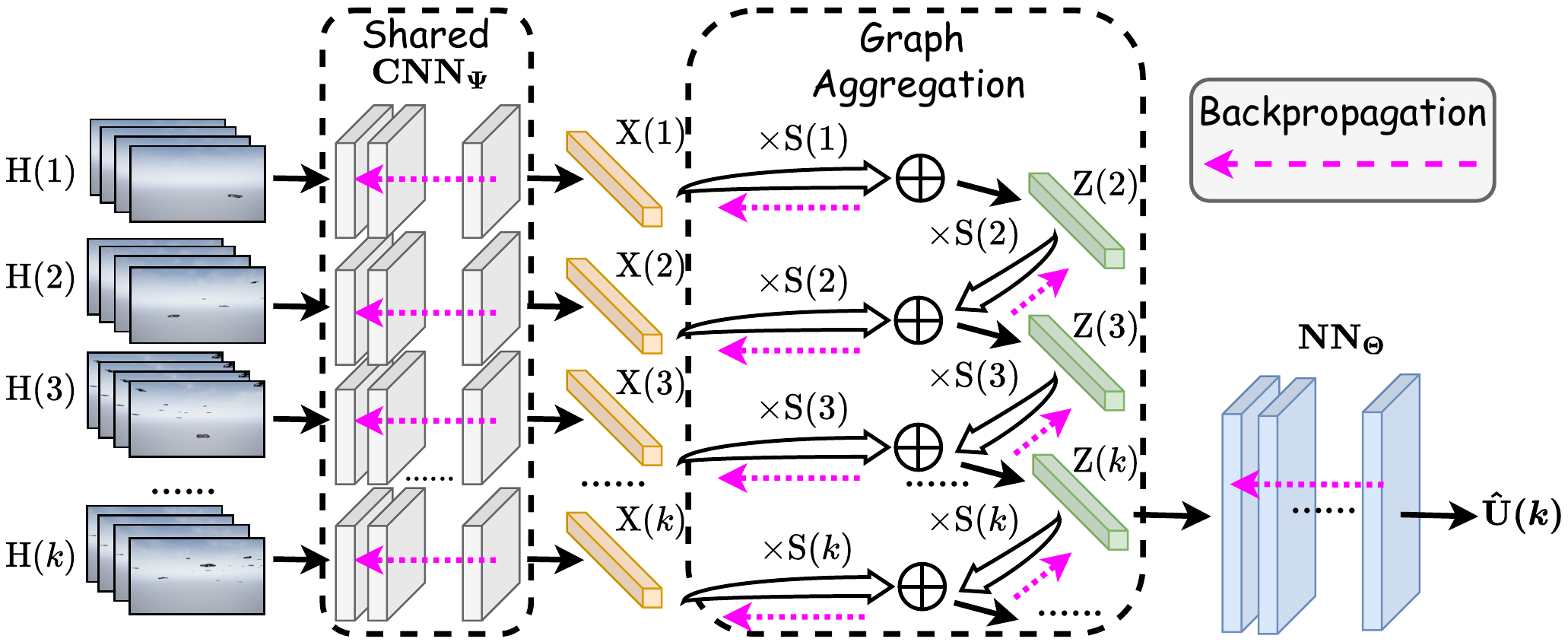}
  \vspace{0.5em}
    \caption{Overview of the end-to-end training of VGAI framework. To obtain the graph sequence $\bbZ(K)$ for the imitation learning of optimal actions $\bbU^{\ast}(K)$, we consecutively execute graph sequence aggregation \eqref{eqn:aggregationSequence} from time step $1$ to $K$ (\ie for $K-1$ steps). Then BPTT is carried out by passing through  $\text{NN}_{\bbTheta}(\cdot)$ once and unrolling $\text{CNN}_{\bbPsi}(\cdot)$ $K$ times.}
    \label{fig:end2end}
   \vspace{-1.5em}
\end{wrapfigure}

Since the local graph aggregation \eqref{eqn:aggregationSequence} is also fully differentiable, 
$\text{CNN}_{\bbPsi}(\cdot)$ can be jointly learned with $\text{NN}_{\bbTheta}(\cdot)$, by end-to-end backward propagation through time (BPTT), as illustrated in Figure~\ref{fig:end2end}. We update the parameters $\{\bbPsi, \bbTheta\}$ by $K$-step BPTT, given the $K$ consecutive GSO matrices $\{\bbS(t)\}_{t=1}^{K}$ and images $\{\bbH(t)\}_{t=1}^{K}$.


\subsection{Implementation Details}
Our default scenario considers a large flock of $N = 50$ agents with a communication radius of $R = 1.5\text{m}$ and a discretization time period of $T_{s} = 0.01\text{s}$. 

The flock locations were initialized uniformly on the disc with radius $\sqrt{N}$ to normalize the density of agents for changing flock sizes. Initial agent velocities are controlled by a parameter $v_{init} = 3.0 \text{m}/\text{s}$. Agent velocities are sampled uniformly from the interval $[-v_{\text{init}},+v_{\text{init}}]$ and then a bias for the whole flock is added, also sampled from $[-\frac{3}{10} v_{\text{init}},+\frac{3}{10}  v_{\text{init}}]$. To eliminate unsolvable cases, configurations are resampled if any agent fails to have at least two neighbors or if agents begin closer than $0.2 \text{m}$. Finally, acceleration commands are saturated to the range $[-30, 30] \text{m}/\text{s}^{2}$ to improve the numerical stability of training.

The controller was trained to predict the best action class with a $L1$ loss and the optimizer is Adam with a learning rate $0.001$. We also adopted the Dataset Aggregation (DAGger) \citep{DAG2010} algorithm by following the learner's policy rather than the expert's with a probability $0.33$, to account for the inconsistency of training and testing phase state distributions. 

\begin{figure}[ht]
    \centering
    \includegraphics[width=1\linewidth]{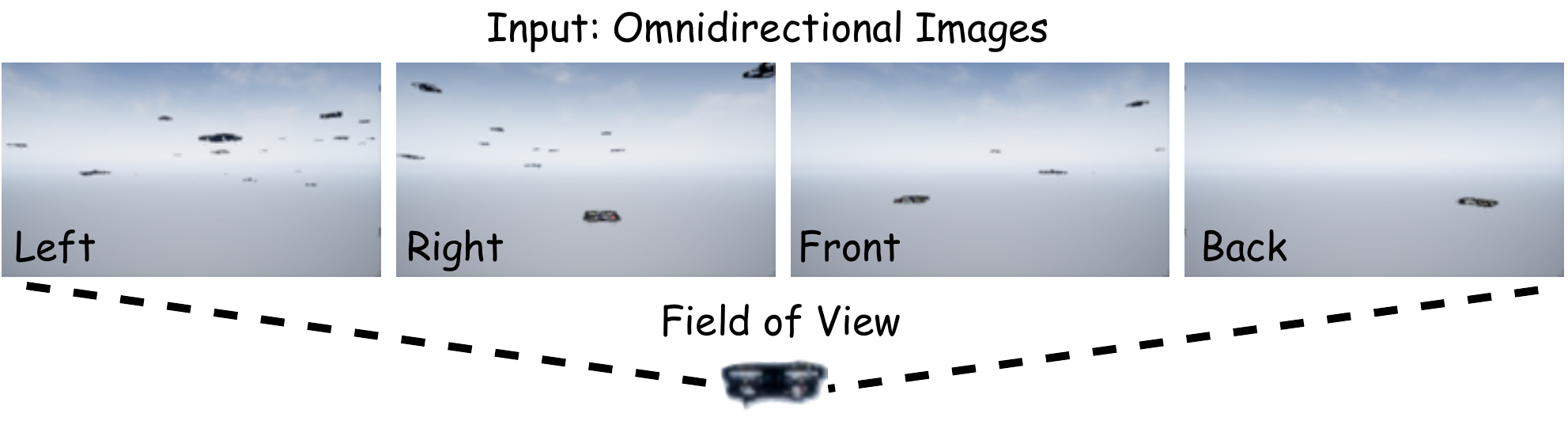}
    \vspace{-0.5em}
    \caption{An example of the rendered raw visual input from the Microsoft Airsim Simulation  environments. The simulation environment allows the user to render four camera-view images, \ie front-center, front-left, front-right, and back. Each rendered view has $144 \times 256$ resolution. 
    At each time step, we concatenate the four views to obtain an agent's complete field of view, which is fed for its visual state feature learning. For the camera configuration, we adjust the quaternion orientation of front-left and front-right cameras to $-0.82$ and $0.82$.
    We also set the aptitude of each agent in-between $37$\text{m} and $43$\text{m} for collision-free random initialization. 
    }
    \vspace{-1em}
    \label{fig:vis-input}
\end{figure}

We conducted the experiments using the Microsoft Airsim Simulation environment, that can render visual inputs for agents. Figure \ref{fig:vis-input} displays a visual observation snapshot for one agent during testing.

\vspace{-2.0em}
\section{Experimental Results}

In this section, we simulate the flocking problem \eqref{eqn:flockingObjective} and evaluate the VGAI controller. We first note that VGAI is the only approach so far that integrates visual observations and communication; moreover, there was no previous success reported on learning decentralized controller over large swarms from  solely visual observations. Therefore, we instead compare the VGAI controller with:
\begin{itemize}
    \item the centralized controller \eqref{eqn:optimalSolution}, that offers an empirical \textbf{lower bound} (here the lower the better) on VGAI performance, since no decentralized controller is supposed to outperform the centralized one. Therefore in tables hereafter, we report velocity variation costs \eqref{eqn:flockingObjective} \textit{normalized} to that of the centralized controller,  for all decentralized/local controllers. 
    \item the DAGNN acting directly on precise neighboring velocity features, following the identical implementation of \cite{Tolstaya19-Flocking}. This offers another \textbf{lower bound} on VGAI's expected performance, since it acts directly on the (ideal) exact knowledge of the relative position and velocity of neighbors, while VGAI requires to extract information from visual inputs.
    \item the local heuristic controller \eqref{eqn:localHeuristic}, which we expect to be an \textbf{upper bound} for VGAI. 
\end{itemize}
In any case, we note from exhaustive simulations, that relative velocity variations less than $3$ would in general achieve successful flocking of a team.



We ran three sets of experiments. First, in Section~\ref{subsec:hyperparams}, we test different hyperparameter selections. Second, in Section~\ref{subsec:initial}, we test different initial conditions of the flocking to assess the robustness of the proposed method. Third, in Section~\ref{subsec:generalization}, we ran a scalability experiment, to show how the proposed method, trained on a given number of agents, still works effectively when tested on larger teams. We train all architectures on teams of $N=50$ agents. We conducted the experiments on the Microsoft Airsim Simulation environment and the detail camera configuration could be referred in Fig.~\ref{fig:vis-input}. 


\subsection{Analysis of the selection of hyperparameters} \label{subsec:hyperparams}

{\footnotesize
\begin{table}[ht]
\centering
\caption{\small{Change in relative cost for various choices of hyperparameters. Cost of (i) centralized controller \eqref{eqn:optimalSolution}: $1.00$, (iii) local heuristic \eqref{eqn:localHeuristic}: $4.13$. (\subref{tab:dim-f}) Change in number of features $F$, we note that VGAI works better for larger $F$. (\subref{tab:dim-k}) Change in number of communication exchanges $K-1$, we see that both DAGNN and VGAI improve when more communication exchanges are involved, we also observe that VGAI for just one communication does not achieve successful flocking.}}
\begin{subtable}{0.45\textwidth} 
{\footnotesize
    \centering
    \begin{tabular}{c|c|c|c}
    \toprule
    $F$   & $6$   & $12$  & $24$  \\ \midrule
    DAGNN & $1.70$ & ---   & ---   \\ \midrule
    VGAI  & $2.37$ & $2.38$ & $2.26$ \\ \bottomrule
    \end{tabular}
    \caption{Feature dimension $F$}
    \label{tab:dim-f}
}
\end{subtable}
\hfill
\begin{subtable}{0.45\textwidth}
{\footnotesize
    \centering
    \begin{tabular}{c|c|c|c}
    \toprule
    $K-1$   & $1$   & $2$  & $3$  \\ \midrule
    DAGNN & $1.94$ & $1.82$ & $1.70$ \\ \midrule
    VGAI  & $2.48$ & $2.29$ & $2.26$ \\     \bottomrule
    \end{tabular}
    \caption{Number of exchanges $K-1$}
    \label{tab:dim-k}
}
\end{subtable}
\label{tab:hyperparam}
\vspace{-0.5em}
\end{table}
}

\myparagraph{Visual state feature dimension $F$.} We set $K-1=3$ and increase the dimension $F$ of the feature $\bbX(t)$ extracted from the visual estimation process, increasing the representation power of the decentralized controller. However, this comes at the expense of communication bandwidth, since more values need to be transmitted to neighboring agents. From the results in Table \ref{tab:dim-f}, we observe that the VGAI performance improves as more features are considered. We also note that even for $F=6$ (same number of features as DAGNN), VGAI still achieves successful flocking; meanwhile, the local heuristic \eqref{eqn:localHeuristic} fails to achieve successful flocking under the same feature dimension.

\myparagraph{Number of communication exchanges $K-1$.} We set $F=24$ and change the number of communication exchanges, which determine how far information can be gathered. Although this information is delayed [cf. \eqref{eqn:aggregationSequence}], it can still be useful to consider it. Note that this comes at the expense of increased communication cost. In Table~\ref{tab:dim-k} we observe that both the DAGNN and the VGAI improve as more communication exchanges are carried out. This suggests that encouraging more communication exchanges can potentially improve flocking behaviors. 

\subsection{Analysis of the initial conditions} \label{subsec:initial}

{\footnotesize
\begin{table}[ht]
\centering
\caption{\small{Change in relative cost for various choices of initial conditions. Cost of (i) centralized controller \eqref{eqn:optimalSolution}: $1.0$ (\subref{tab:init-v}) Maximum initial velocity $v_{\text{init}}$, we see that performance degraded when the agents start flying at higher speeds. (\subref{tab:init-R}) Communication radius $R$, we observe that the performance improve for larger  $R$.}}
\begin{subtable}{0.45\textwidth}
{\footnotesize
    \centering
    \begin{tabular}{c|c|c|c}
    \toprule
    $v_{\text{init}}$ [$\text{m}/\text{s}$]   & $1$   & $2$  & $3$  \\ \midrule
    DAGNN & $1.58$  & $1.63$  & $1.70$  \\ \midrule
    Local & $3.63$ & $3.77$ & $4.13$ \\ \midrule
    VGAI  & $1.97$  & $2.10$  & $2.26$  \\ \bottomrule
    \end{tabular}
    \caption{Initial velocity $v_{\text{init}}$}
    \label{tab:init-v}
}
\end{subtable}
\hfill
\begin{subtable}{0.45\textwidth}
{\footnotesize
    \centering
    \begin{tabular}{c|c|c|c}
    \toprule
    $R$ [$\text{m}$]   & $1.0$   & $1.5$  & $2.0$  \\ \midrule
    DAGNN & $1.95$  & $1.70$  & $1.59$  \\ \midrule
    Local & $4.85$  & $4.13$ & $2.75$ \\ \midrule
    VGAI  & $2.65$  & $2.26$  & $1.77$  \\ \bottomrule
    \end{tabular}
    \caption{Communication radius $R$}
    \label{tab:init-R}
}
\end{subtable}
\label{tab:initial}
\vspace{-0.5em}
\end{table}
}

\myparagraph{Maximum Initial Velocity $v_{\text{init}}$.} We set $F=24$ and $K-1=3$ based on Section~\ref{subsec:hyperparams}, and set $R=1.5\text{m}$. Increasing maximum initial velocity causes the agents to start randomly moving in different directions with higher speeds. This makes the controllability harder, since there is less time to rein in the agents, before they get disconnected from the communication network. We observe from the results in Table~\ref{tab:init-v} that this is indeed the case, since the performance of all controllers gets degraded for increasing $v_{\text{init}}$. 

\myparagraph{Communication Radius $R$.} We set $F=24$ and $K-1=3$ based on Section~\ref{subsec:hyperparams}, and set $v_{\text{init}}=3\text{m}/\text{s}$. The communication radius determines the range within which each agent can communicate directly with other agents, i.e. all agents located within the communication radius are considered to be neighbors at that specific time instant. Thus, the communication radius determines the degree of the communication network. Increasing the communication radius implies that each agent can communicate with a larger number of agents and receive further away information without increased delays. Thus, increasing the communication radius should improve the performance of the decentralized controllers. We see in Table~\ref{tab:init-R} that this is, indeed, the case. Interestingly, we note that for $R=2.0\text{m}$ the performance of the VGAI almost matches that of the DAGNN.



\subsection{Scalability to larger teams: Generalization} \label{subsec:generalization}

\begin{wraptable}{r}{60mm}
\vspace{-1.5em}
\centering
\caption{\small{Scalablity to larger teams. VGAI controller is trained on a team of $N=50$ agents, and then tested on larger teams $N'$. This experiment shows the capability of the VGAI to transfer at scale, that is, to be trained in smaller teams but be deployed in larger ones.}}
\begin{tabular}{c|c|c|c}
\toprule
$N'$ & 50    & 60    & 75    \\ \midrule
VGAI & $1.70$ & $2.14$ & $2.23$ \\ \bottomrule
\end{tabular}
\vspace{-0.5em}
\label{tab:generality}
\end{wraptable}

So far, we have studied the dependence of the VGAI controller with different values of hyperparameters (Section~\ref{subsec:hyperparams}) and found out that more features and more exchanges improve the controller, albeit at the expense of increased communication cost. Then, we studied the robustness of the VGAI controller to different initial conditions (Section~\ref{subsec:initial}) and found that higher initial speeds make the control problem harder, while a larger communication radius makes it easier. In what follows, we set $F=24$, $K-1=3$, $v_{\text{init}}=3\text{m}/\text{s}$ and $R=1.5\text{m}$, and study the scalability of the VGAI controller.

We train the controller for a team of $N=50$ agents, and then we test it for teams of increasing size $N'$. We note that, since the agents only rely on neighboring communications [cf. \eqref{eqn:aggregationSequence}], and since the learned parameters are shared across all agents [cf. \eqref{eqn:NN}], the VGAI controller is agnostic to the team size. This is due to the permutation equivariace and stability properties of the GNN \cite{Gama20-Stability}. We see in Table~\ref{tab:generality} that a VGAI based controller trained on $N=50$ agents can successfully flock a team of agents of up to $N'=75$ agents. This suggests that the VGAI controller can be trained on smaller networks and stil be successfully deployed on larger ones.

\vspace{-1.0em}
\section{Conclusion}
\vspace{-1.0em}
This paper presented a vision-based decentralized controller learning (VGAI) framework for large-scale robot swarms. We demonstrated the feasibility of CNN-GNN cascaded networks for automatically learning decentralized
controllers. It can operate with large teams of agents based only on local visual observation, with coupled state dynamics and sparse communication links. Experimental results quantitatively confirm the value of local neighborhood information to the stability of controlled flocks. We also show that our learned controller is robust to changes in the range of communication radius, number of agents and maximum initial speed of the flock. 
Our future work will further improve the cohesion behaviours by exploring other end-to-end visual learning frameworks.



\clearpage


\bibliography{Drone}  

\begin{thebibliography}{32}
\providecommand{\natexlab}[1]{#1}
\providecommand{\url}[1]{\texttt{#1}}
\expandafter\ifx\csname urlstyle\endcsname\relax
  \providecommand{\doi}[1]{doi: #1}\else
  \providecommand{\doi}{doi: \begingroup \urlstyle{rm}\Url}\fi

\bibitem[{Mellinger} and {Kumar}(2011)]{position2011}
D.~{Mellinger} and V.~{Kumar}.
\newblock Minimum snap trajectory generation and control for quadrotors.
\newblock In \emph{ICRA}, 2011.

\bibitem[Kushleyev et~al.(2013)Kushleyev, Mellinger, Powers, and
  Kumar]{Kushleyev2013}
A.~Kushleyev, D.~Mellinger, C.~Powers, and V.~Kumar.
\newblock Towards a swarm of agile micro quadrotors.
\newblock \emph{Autonomous Robots}, 2013.

\bibitem[{Preiss} et~al.(2017){Preiss}, {Honig}, {Sukhatme}, and
  {Ayanian}]{preiss2017}
J.~A. {Preiss}, W.~{Honig}, G.~S. {Sukhatme}, and N.~{Ayanian}.
\newblock Crazyswarm: A large nano-quadcopter swarm.
\newblock In \emph{ICRA}, 2017.

\bibitem[{Weinstein} et~al.(2018){Weinstein}, {Cho}, {Loianno}, and
  {Kumar}]{weinstein2018}
A.~{Weinstein}, A.~{Cho}, G.~{Loianno}, and V.~{Kumar}.
\newblock Visual inertial odometry swarm: An autonomous swarm of vision-based
  quadrotors.
\newblock \emph{IEEE RA-L}, 2018.

\bibitem[V{\'{a}}s{\'{a}}rhelyi et~al.(2014)V{\'{a}}s{\'{a}}rhelyi,
  Vir{\'{a}}gh, Somorjai, Tarcai, Sz{\"{o}}r{\'{e}}nyi, Nepusz, and
  Vicsek]{vasarhelyiVSTSNV2014}
G.~V{\'{a}}s{\'{a}}rhelyi, C.~Vir{\'{a}}gh, G.~Somorjai, N.~Tarcai,
  T.~Sz{\"{o}}r{\'{e}}nyi, T.~Nepusz, and T.~Vicsek.
\newblock Outdoor flocking and formation flight with autonomous aerial robots.
\newblock \emph{CoRR}, 2014.

\bibitem[V{\'a}s{\'a}rhelyi et~al.(2018)V{\'a}s{\'a}rhelyi, Vir{\'a}gh,
  Somorjai, Nepusz, Eiben, and Vicsek]{varhelyiea2018}
G.~V{\'a}s{\'a}rhelyi, C.~Vir{\'a}gh, G.~Somorjai, T.~Nepusz, A.~E. Eiben, and
  T.~Vicsek.
\newblock Optimized flocking of autonomous drones in confined environments.
\newblock \emph{Science Robotics}, 2018.

\bibitem[Reynolds(1987)]{global1987}
C.~W. Reynolds.
\newblock Flocks, herds and schools: A distributed behavioral model.
\newblock In \emph{ACM SIGGRAPH}, 1987.

\bibitem[Tanner et~al.(2003)Tanner, Jadbabaie, and Pappas]{global2003}
H.~Tanner, A.~Jadbabaie, and G.~Pappas.
\newblock Stable flocking of mobile agents, part ii: Dynamic topology.
\newblock \emph{Departmental Papers (ESE)}, 2003.

\bibitem[Floreano and Wood(2015)]{swarm2015}
D.~Floreano and R.~Wood.
\newblock Science, technology and the future of small autonomous drones.
\newblock \emph{Nature}, 2015.

\bibitem[Tolstaya et~al.(2019)Tolstaya, Gama, Paulos, Pappas, Kumar, and
  Ribeiro]{Tolstaya19-Flocking}
E.~Tolstaya, F.~Gama, J.~Paulos, G.~Pappas, V.~Kumar, and A.~Ribeiro.
\newblock Learning decentralized controllers for robot swarms with graph neural
  networks.
\newblock In \emph{CoRL}, 2019.

\bibitem[Zhu et~al.(2018)Zhu, Wen, Du, Bian, Ling, Hu, Nie, Cheng, Liu, Liu,
  et~al.]{zhu2018visdrone}
P.~Zhu, L.~Wen, D.~Du, X.~Bian, H.~Ling, Q.~Hu, Q.~Nie, H.~Cheng, C.~Liu,
  X.~Liu, et~al.
\newblock Visdrone-det2018: The vision meets drone object detection in image
  challenge results.
\newblock In \emph{Proceedings of the European Conference on Computer Vision
  (ECCV) Workshops}, 2018.

\bibitem[Wu et~al.(2019)Wu, Suresh, Narayanan, Xu, Kwon, and
  Wang]{wu2019delving}
Z.~Wu, K.~Suresh, P.~Narayanan, H.~Xu, H.~Kwon, and Z.~Wang.
\newblock Delving into robust object detection from unmanned aerial vehicles: A
  deep nuisance disentanglement approach.
\newblock In \emph{Proceedings of the IEEE International Conference on Computer
  Vision}, pages 1201--1210, 2019.

\bibitem[Schilling et~al.(2018)Schilling, Lecoeur, Schiano, and
  Floreano]{schilling2018learning}
F.~Schilling, J.~Lecoeur, F.~Schiano, and D.~Floreano.
\newblock Learning vision-based cohesive flight in drone swarms.
\newblock \emph{IEEE RA-L}, 2018.

\bibitem[Gama et~al.(2019{\natexlab{a}})Gama, G.~Marques, Leus, and
  Ribeiro]{Gama_2019}
F.~Gama, A.~G.~Marques, G.~Leus, and A.~Ribeiro.
\newblock Convolutional neural network architectures for signals supported on
  graphs.
\newblock \emph{IEEE TSP}, 2019{\natexlab{a}}.

\bibitem[Gama et~al.(2019{\natexlab{b}})Gama, G.~Marques, Ribeiro, and
  Leus]{Gama19-Aggregation}
F.~Gama, A.~G.~Marques, A.~Ribeiro, and G.~Leus.
\newblock Aggregation graph neural networks.
\newblock In \emph{ICASSP}, 2019{\natexlab{b}}.

\bibitem[Witsenhausen(1968)]{witsenhausen1968counterexample}
H.~S. Witsenhausen.
\newblock A counterexample in stochastic optimum control.
\newblock \emph{SICON}, 1968.

\bibitem[{Jadbabaie} et~al.(2003){Jadbabaie}, {Jie Lin}, and
  {Morse}]{local2003}
A.~{Jadbabaie}, {Jie Lin}, and A.~S. {Morse}.
\newblock Coordination of groups of mobile autonomous agents using nearest
  neighbor rules.
\newblock \emph{IEEE TACON}, 2003.

\bibitem[Ross et~al.(2012)Ross, Melik{-}Barkhudarov, Shankar, Wendel, Dey,
  Bagnell, and Hebert]{mav2012}
S.~Ross, N.~Melik{-}Barkhudarov, K.~S. Shankar, A.~Wendel, D.~Dey, J.~A.
  Bagnell, and M.~Hebert.
\newblock Learning monocular reactive {UAV} control in cluttered natural
  environments.
\newblock \emph{CoRR}, 2012.

\bibitem[{Loquercio} et~al.(2018){Loquercio}, {Maqueda}, {del-Blanco}, and
  {Scaramuzza}]{dronet2018}
A.~{Loquercio}, A.~I. {Maqueda}, C.~R. {del-Blanco}, and D.~{Scaramuzza}.
\newblock Dronet: Learning to fly by driving.
\newblock \emph{IEEE RA-L}, 2018.

\bibitem[Sadeghi and Levine(2016)]{rf2016}
F.~Sadeghi and S.~Levine.
\newblock (cad){\textdollar}{\^{}}2{\textdollar}rl: Real single-image flight
  without a single real image.
\newblock \emph{CoRR}, 2016.

\bibitem[{Giusti} et~al.(2016){Giusti}, {Guzzi}, {Cireşan}, {He},
  {Rodríguez}, {Fontana}, {Faessler}, {Forster}, {Schmidhuber}, {Caro},
  {Scaramuzza}, and {Gambardella}]{vision2016}
A.~{Giusti}, J.~{Guzzi}, D.~C. {Cireşan}, F.~{He}, J.~P. {Rodríguez},
  F.~{Fontana}, M.~{Faessler}, C.~{Forster}, J.~{Schmidhuber}, G.~D. {Caro},
  D.~{Scaramuzza}, and L.~M. {Gambardella}.
\newblock A machine learning approach to visual perception of forest trails for
  mobile robots.
\newblock \emph{IEEE RA-L}, 2016.

\bibitem[Gandhi et~al.(2017)Gandhi, Pinto, and Gupta]{vision2017}
D.~Gandhi, L.~Pinto, and A.~Gupta.
\newblock Learning to fly by crashing.
\newblock \emph{CoRR}, 2017.

\bibitem[Smolyanskiy et~al.(2017)Smolyanskiy, Kamenev, Smith, and
  Birchfield]{vision2017_2}
N.~Smolyanskiy, A.~Kamenev, J.~Smith, and S.~Birchfield.
\newblock Toward low-flying autonomous {MAV} trail navigation using deep neural
  networks for environmental awareness.
\newblock \emph{CoRR}, 2017.

\bibitem[{Faigl} et~al.(2013){Faigl}, {Krajník}, {Chudoba}, {Přeučil}, and
  {Saska}]{marker2013}
J.~{Faigl}, T.~{Krajník}, J.~{Chudoba}, L.~{Přeučil}, and M.~{Saska}.
\newblock Low-cost embedded system for relative localization in robotic swarms.
\newblock In \emph{ICRA}, 2013.

\bibitem[Krajník et~al.(2014)Krajník, Nitsche, Faigl, Vaněk, Saska, Duckett,
  and Mejail]{marker2014}
T.~Krajník, M.~Nitsche, J.~Faigl, P.~Vaněk, M.~Saska, T.~Duckett, and
  M.~Mejail.
\newblock \emph{JINT}, 2014.

\bibitem[Sandryhaila and Moura(2013)]{Sandryhaila13-DSPG}
A.~Sandryhaila and J.~M.~F. Moura.
\newblock Discrete signal processing on graphs.
\newblock \emph{IEEE TSP}, 2013.

\bibitem[Shuman et~al.(2013)Shuman, Narang, Frossard, Ortega, and
  Vandergheynst]{Shuman13-SPG}
D.~I. Shuman, S.~K. Narang, P.~Frossard, A.~Ortega, and P.~Vandergheynst.
\newblock The emerging field of signal processing on graphs: Extending
  high-dimensional data analysis to networks and other irregular domains.
\newblock \emph{IEEE SPM}, 2013.

\bibitem[Ortega et~al.(2018)Ortega, Frossard, Kova{\v{c}}evi{\'{c}}, Moura, and
  Vandergheynst]{Ortega18-GSP}
A.~Ortega, P.~Frossard, J.~Kova{\v{c}}evi{\'{c}}, J.~M.~F. Moura, and
  P.~Vandergheynst.
\newblock Graph signal processing: Overview, challenges and applications.
\newblock \emph{PIEEE}, 2018.

\bibitem[Goodfellow et~al.(2016)Goodfellow, Bengio, and
  Courville]{Goodfellow16-DeepLearning}
I.~Goodfellow, Y.~Bengio, and A.~Courville.
\newblock \emph{Deep Learning}.
\newblock The {MIT} Press, 2016.

\bibitem[Ross and Bagnell(2010)]{Ross10-ImitationLearning}
S.~Ross and D.~Bagnell.
\newblock Efficient reductions for imitation learning.
\newblock In \emph{AISTATS}, 2010.

\bibitem[Ross et~al.(2010)Ross, Gordon, and Bagnell]{DAG2010}
S.~Ross, G.~J. Gordon, and J.~A. Bagnell.
\newblock No-regret reductions for imitation learning and structured
  prediction.
\newblock \emph{CoRR}, 2010.

\bibitem[Gama et~al.(2020)Gama, Bruna, and Ribeiro]{Gama20-Stability}
F.~Gama, J.~Bruna, and A.~Ribeiro.
\newblock Stability properties of graph neural networks.
\newblock \emph{arXiv:1905.04497v4 [cs.LG]}, 8 July 2020.
\newblock URL \url{http://arxiv.org/abs/1905.04497}.

\end{thebibliography}

\end{document}